# Ferromagnetic Semiconductors: Moving Beyond (Ga,Mn)As

A. H. MacDonald,[1] P. Schiffer,[2] and N. Samarth[2]

[1]Department of Physics. University of Texas at Austin, Austin, Texas 78712 USA

[2]Department of Physics and Materials Research Institute, Pennsylvania State University, University Park, Pennsylvania 16802 USA

The recent development of MBE techniques for growth of III-V ferromagnetic semiconductors has created materials with exceptional promise in *spintronics*, i.e. electronics that exploit carrier spin polarization. Among the most carefully studied of these materials is (Ga,Mn)As, in which meticulous optimization of growth techniques has led to reproducible materials properties and ferromagnetic transition temperatures well above 150 K. We review progress in the understanding of this particular material and efforts to address ferromagnetic semiconductors as a class. We then discuss proposals for how these materials might find applications in spintronics. Finally, we propose criteria that can be used to judge the potential utility of newly discovered ferromagnetic semiconductors, and we suggest guidelines that may be helpful in shaping the search for the ideal material.

## 1. Background and Introduction

According to legend, Wolfgang Pauli offered some infamously bad advice regarding the sensitivity of semiconductor properties to impurities. ("One shouldn't work on semiconductors, that is a filthy mess; who knows whether any semiconductors exist."[1]) Some 70 years later, we now possess an elaborate technology that makes exquisite use of those formerly troublesome impurities. Researchers are attempting to add a sequel to this happy story by integrating magnetism into these vital material systems.[2,3,4] *Semiconductor spintronics* could enable seamless electrical manipulation of magnetic states and magnetic modification of electrical signals. One can envision, for example, devices like magnetic bipolar transistors[5] transferring information from magnetic media to conventional semiconductor circuits. An important step toward this technology is to find semiconducting materials which display electrically tunable ferromagnetism at temperatures well above room temperature *and* which can be incorporated into complex integrated circuits. The list of candidate materials is expanding rapidly (see references 6,7,8,9,10,11,12,13,14,15,16,17,18,19,20 and work cited therein), and it is difficult to guess which is most likely to satisfy *both* criteria. Substantial progress has been made, however, in understanding the materials whose remarkable properties initiated this quest: $Ga_{1-x}Mn_xAs$ and $In_{1-x}Mn_xAs$. Ferromagnetic critical temperature aside (currently $T_c <$ 170 K for $Ga_{1-x}Mn_xAs$ and lower for $In_{1-x}Mn_xAs$) already comes close to realizing the suite of coupled and tunable magnetic, transport, and optical properties that is imagined in many spintronic device concepts. For instance, electrically tunable ferromagnetism has been explicitly demonstrated in $In_{1-x}Mn_xAs$-based field effect devices.[21]

Ferromagnetic semiconductors are of course not new,[22] and carrier-mediated ferromagnetism, the handle that allows electrical control, was also demonstrated some time ago[23] although only at liquid helium temperatures. Interest in ferromagnetic semiconductors was invigorated several years ago, however, by the discovery of ferromagnetism at relatively high temperatures ($T_C > 100$ K) in diluted magnetic semiconductors (DMS) derived from technologically important III-V compounds InAs[24,25] and GaAs.[26] This discovery struck an immediate chord with researchers interested in spintronic applications, since the *host* semiconductors in these materials are well understood and already play a crucial technological role.

As we will discuss below, we cannot be certain that $Ga_xMn_{1-x}As$ or $In_xMn_{1-x}As$ will ever reach the high transition temperatures required for widespread applications. Nonetheless, the understanding of these two materials is already at a stage where they can provide a test bed for simple spintronic device concepts and functionalities. We argue here that the progress made with these materials suggests strategies that can help guide the search for new DMS ferromagnets. We focus on $Ga_{1-x}Mn_xAs$, in which demonstrably carrier-mediated ferromagnetism already persists up to rather high temperatures[27, 28, 29, 30]. We first address recent advances in the experimental materials science of $Ga_{1-x}Mn_xAs$, illustrating how the understanding and control of defects is essential to exploiting the potential of these materials. We then review theoretical progress, showing how even idealized models can provide a usefully predictive understanding of these complex materials. Next, we discuss progress and prospects for devices incorporating these materials, and we speculate on the materials requirements for ferromagnetic semiconductors to have widespread technological impact. Finally, we conclude with some further speculations on what kinds of materials are most promising in the quest for the ideal DMS ferromagnet.

**2. Recent progress in the materials science of $Ga_{1-x}Mn_xAs$**

When Mn is substituted for Ga in a GaAs lattice (see Fig. 1), it acts as an acceptor, providing holes that mediate a ferromagnetic interaction between the local moments of the open d shells in the Mn atoms. This picture has been established experimentally by the strong correlation between the ferromagnetic $T_C$ and the carrier concentration demonstrated in Fig. 1. Collective ferromagnetic order requires that a minimum of about 2% of the Ga ions be replaced on the lattice by Mn ions, in order to provide a sufficiently high density of carriers. To incorporate such a large concentration of Mn in the GaAs lattice without forming inclusions of the thermodynamically more stable metallic MnAs phase, $Ga_{1-x}Mn_xAs$ must be grown by molecular beam epitaxy at relatively low temperatures ($T_{substrate} \sim 250°$ C). Even during the early stages of study, it was recognized that this low growth temperature results in a high density of point defects. The most important of these are now known to be As antisites and Mn interstitials[31] that act as double donors, compensating a significant fraction of the free holes.

Early studies suggested that defects limited the ferromagnetic transition temperature $T_c$ to a maximum value of $T_c \sim 110$ K for $x \sim 0.05$.[32] These samples often had relatively unconventional magnetic properties, including an unusual temperature dependence of the magnetization, ascribed to the existence of multiple exchange constants[33] or to noncollinear order.[34] This emerging picture of $Ga_{1-x}Mn_xAs$ was unsettled by the startling discovery[35] that some of the compensating defects were energetically metastable, and post-growth annealing at relatively low temperatures could dramatically modify key sample parameters like the lattice constant, the hole density, and the ferromagnetic transition temperature (see Fig. 2).[36] These initial annealing experiments yielded three very significant results: the Curie temperature of as-grown $Ga_{1-x}Mn_xAs$ samples could be reproducibly increased to ~110 K for a wide range of sample compositions; annealing transformed the temperature dependence of the magnetization into the form expected for homogeneous collinearly ordered ferromagnets;[37] and systematic Rutherford backscattering and particle-induced x-ray emission (PIXE) experiments strongly suggested that the annealing process reduced the population of Mn interstitials. Further studies at several institutions showed that the Curie temperature of annealed $Ga_{1-x}Mn_xAs$ *could* exceed 110 K,[28-30] reaching record values as high as ~170 K in carefully engineered heterostructures. The key appears to lie in the proximity and nature of the surface of the $Ga_{1-x}Mn_xAs$ layer:[29,30] for a fixed set of annealing conditions (e.g. duration and temperature), the maximum attainable Curie temperature decreases with increasing sample thickness. Furthermore, the Curie temperature is also limited by capping the $Ga_{1-x}Mn_xAs$ layer, even with a few nanometers of epitaxial GaAs.[38] We now understand that this thickness and capping behavior is related to the diffusion of interstitial Mn: since the interstitials are donors, they compensate the free holes provided by the substitutional Mn ions (i.e. those on the Ga lattice sites). Thermal annealing drives the Mn interstitials to the free surface, increasing the free hole concentration and correspondingly the ferromagnetic transition temperature.[39]

This demonstration of the crucial importance of defects, thermal treatments and neighboring layers has significant implications for device applications and for the fundamental understanding of ferromagnetic semiconductors. In particular, it suggests that one can almost always tweak transition temperatures upward in other ferromagnetic semiconductors by altering key defects. Recent experiments have also probed numerous other aspects of the fundamental physics of $Ga_{1-x}Mn_xAs$ such as the magnetic anisotropy,[40,41] dc and ac conductivity[42,43,44] the band structure[45,46] and the spin polarization,[47] leading to a rather thorough experimental understanding.[48] Several possible strategies are currently being explored to push $T_C$ still higher in $Ga_{1-x}Mn_xAs$ or $In_{1-x}Mn_xAs$. These include co-doping with other materials to increase the free-carrier density, and wave function engineering in heterojunction systems to increase the effectiveness of the exchange interactions. It seems far from certain, however, that these efforts will yield ferromagnetism at high enough temperature for applications, motivating the study of other related materials discussed below.

**3. Theoretical Pictures**

Experiments have established that ferromagnetism in $Ga_{1-x}Mn_xAs$ is mediated by a rather low density of valence band holes, the key property that allows magnetic properties to be altered electrically. This occurrence of carrier-mediated ferromagnetism at reasonably high temperatures in a magnetically and electrically dilute system is quite unusual. Crucially, in $Ga_{1-x}Mn_xAs$ and $In_{1-x}Mn_xAs$ the Mn d electrons have weak valence fluctuations and are not strongly incorporated into the bonding orbitals of the semiconductor. Electrons in the half filled d-shell of $Mn^{++}$ form a quantum state with spin $S=5/2$. According to the third law of thermodynamics, the macroscopic entropy associated with arbitrary local moment spin orientations must vanish at low temperatures; spins must become fixed or participate in a correlated-fluctuating-moment quantum state. In many (II,Mn)VI semiconductors, the Mn moments interact very weakly with each other unless they happen to lie on neighboring cation cites, and they fluctuate randomly in orientation down to very low temperatures (as illustrated in the upper panel of Fig. 3). The introduction of these low-energy degrees of freedom creates the opportunity for a ferromagnetic state if an energetic preference for moment alignment can be engineered.

In transition-metal based (III,V) ferromagnetic semiconductors, the occurrence of robust ferromagnetism as described in the previous section makes it evident that this coupling is present. It seems clear that the ultimate origin of Mn moment alignment is the hybridization that occurs between d-orbitals on cation sites and orbitals in the partially filled valence band that are centered on neighboring anion sites. Mn substitution introduces both local moments and valence band holes that hybridize with Mn d-orbitals of the same spin. The energy of the system is lowered when the unoccupied levels near the top of the valence band (holes) have the same spin orientation as the Mn d-orbitals, that is when the total valence band spin is opposite to that of the Mn ion. The valence band hole will then tend to align any Mn moments with which it overlaps as suggested by the middle panel of Fig. 3. These interactions follow essentially from quantum mechanical level repulsion and strengthen when the energetic separation between the occupied Mn d levels and the hole states at the top of the valence band gets smaller.

A key issue[49] for the theory of the III-V DMS ferromagnets is whether the carriers end up residing in a tightly-bound anti-bonding state that has primarily d-character, or in a more spatially extended structure that has primarily the p-character of the host valence band. Because the moments in DMS systems are dilute, large values for the inter-moment couplings responsible for carrier-mediated ferromagnetism require acceptor level states that extend over at least a few lattice constants. Strong hybridization strengthens the coupling between Mn and band-spin orientations, but also shortens the range of its impact by localizing the acceptor level. The optimal hybridization strength likely lies somewhere in the middle ground between shallow and deep acceptor levels; this is the space that $Ga_{1-x}Mn_xAs$ appears to occupy.

$Ga_{1-x}Mn_xN$, one of the candidate materials for room temperature DMS ferromagnetism,[14-50] appears to have strongly localized holes,[51,52] suggesting that the microscopic physics of the high-temperature ferromagnetism reported in this material may have a different character. Although we do not have a rigorous argument that would exclude the possibility of such a strongly localized impurity band system attaining ferromagnetic order, it seems clear to us that it would cost relatively little energy to rotate the spin of an impurity-band electron centered on one magnetic site relative to that of an impurity-band electron centered on a nearby magnetic site. Because this energy cost is the stiffness that supports magnetization, $T_C$ in a diluted moment system would tend to be low if ferromagnetism were mediated by an impurity band.

This qualitative theoretical picture of DMS ferromagnetism in $Ga_{1-x}Mn_xAs$ has been substantiated by combined insights from several different approaches. Extensive *ab initio* electronic structure studies of these systems are able to account for important materials trends and qualitative features of the exchange interactions between Mn ions and band electrons.[49,53,54,55,56,57,58,59] For example, they are able to account for the property that Mn acceptor levels are more localized in nitrides then in arsenides or phosphides. *Ab initio* electronic structure calculations have been key to achieving an understanding of materials trends within both II-VI and III-V based Mn DMS families, and in providing useful guidance for other promising families. The prediction of exchange interaction strengths and ferromagnetic transition temperatures in an individual material is however sensitive to approximations made in treating electronic correlations. It is now generally agreed that the local density approximation does not establish the relative placement of the Mn d-levels and the valence band maximum with sufficient accuracy to yield accurate values for the strength of the interaction between Mn and valence band spins. In particular, the local-density-approximation fails to completely eliminate self-interaction effects that are important for localized orbitals, and therefore tends to place the d-bands at too high an energy relative to the valence band. For $Ga_{1-x}Mn_xAs$, this error results in an overestimate of exchange interaction strengths. LDA+U approximations in density functional theory largely eliminate this deficiency,[48,58,59] and are apparently necessary to obtain reasonably accurate results for the magnetic properties of these materials.

A complementary phenomenological approach,[60,61,62,63,64] which introduces a single free constant $J_{pd}$ to characterize the strength of the exchange interaction between Mn and hole spins cannot address material trends, but has had considerable success in explaining many details of the properties of $Ga_{1-x}Mn_xAs$ and similar materials. The resulting model, in which ferromagnetism arises from exchange coupling between local moments and band electrons, is often referred to as the Zener model.[65,66] As applied to $Ga_{1-x}Mn_xAs$, it assumes that the transition energy from $Mn^{+++}$ to $Mn^{++}$ is well below the top of the valence band and (less critically for low-energy properties) that the transition energy from $Mn^{++}$ to $Mn^+$ is well above the bottom of the conduction band. When valid, these assumptions guarantee that $Mn^{++}$ valence fluctuations are virtual and can be represented by an exchange interaction. As applied to $Ga_{1-x}Mn_xAs$ and related materials, the Zener model also assumes that valence band hole states in the DMS are formed from states near the valence band edge of the host semiconductor. The single phenomenological constant $J_{pd}$ then represents the spin-dependent scattering amplitude of band electrons at a Mn site, which is assumed to have negligible dependence on either the initial or the final band state, *i.e.* the range of this interaction is assumed to be small compared to the Fermi wavelength. The applicability of this model to any given material, and the value of the phenomenological parameter $J_{pd}$, must be established either by comparison with experiment or on the basis of reliable *ab-initio* calculations. For example, this model was used in early theoretical work[59] to predict properties of materials like (Zn,Mn)O and (Ga,Mn)N, but later experiments have shown that it is not appropriate for these systems. The advantages of the Zener model are two-fold: first, the value of $J_{pd}$ can be taken from experiment, obviating the substantial difficulty of calculating its value reliably from first principles; second, this approach focuses on the degrees of freedom that are important for magnetism, facilitating the analysis of magnetic, transport, and optical properties, for example by increasing the number of independent Mn spins that can be included in numerical electronic structure and Monte Carlo[67] calculations.

Although the Zener model is simplified, its properties are still complex because of the combined influence of interactions and disorder. It is most easily analyzed when the dilute Mn ions are modeled as a magnetic continuum and spatial fluctuations in the magnetization orientation are not taken into account. The resulting mean-field theory, illustrated in Fig. 4, suggests a strategy that can be used in the search for high $T_c$ materials – look for systems with strong exchange interactions and a large valence band density of states. This tactic likely works only to a point, however, since strong exchange coupling leads to more localized acceptor levels that will not effectively couple Mn ions located at different lattice sites. Similarly, a large valence band density of states implies a large effective mass and this also leads to more localized acceptor levels. Additionally, variations in the sign of strong exchange interactions are expected, on the basis of RKKY arguments, to become more important and lead to more complex magnetic states when the carrier density is too large. These limitations of mean-field theory are summarized schematically in Fig. 4. The fact that the valence band of $Ga_{1-x}Mn_xAs$ has both heavy holes that are easily spin-polarized, opening the door to ferromagnetism, and light holes that are not easily localized expands the range of validity of mean-field-theory and appears to be key to the high transition temperatures. Mean-field theory has been very successful in explaining many magnetic and transport properties of $Ga_{1-x}Mn_xAs$. In mean-field-theory the transition temperature increases without limit with both the exchange interaction strength and with the hole density. These trends cannot continue indefinitely because strong exchange interactions will eventually localize holes and high densities will eventually lead to interactions that vary rapidly in space. Monte Carlo calculations and qualitative arguments[65] suggest that the highest transition temperatures tend to occur at the borders of the mean-field-theory validity region.

Although not highly relevant to applications, $Ga_{1-x}Mn_xAs$ at small x and low hole density is very interesting from a theoretical point of view. In this limit, a single-hole can orient a number of Mn moments, by creating a magnetic polaron. Disorder is critically important, and ferromagnetism, when it occurs, is more fragile. Theorists have also used model approaches with varying degrees of realism to look at this regime, which cannot be described by the simple mean-field theory, accounting for the effects of disorder and localization,[68,69,70,71,72] the possible role of impurity band formation,[73,74] and the role of specific defects.[75,76] This combination of several different approaches has been useful in building an understanding across the entire accessible range of electrical and magnetic doping.

## 4. Ferromagnetic semiconductor heterostructure devices

Semiconductor heterostructure devices (like quantum well lasers, high electron mobility transistors and resonant tunneling diodes) are typically fabricated in a "vertically" integrated fashion, in which the material is compositionally modulated along the axis of epitaxy. The incorporation of ferromagnetic layers of $Ga_{1-x}Mn_xAs$ into such devices sets an important and awkward constraint: once a ferromagnetic $Ga_{1-x}Mn_xAs$ layer has been grown, the rest of the device also must be grown at substrate temperatures of around 250 C or less. Otherwise, the homogeneous ferromagnetic semiconductor "puddles" up into an inhomogeneous, random assembly of ferromagnetic metallic clusters (MnAs) within a GaAs matrix (Such "hybrid" metal-semiconductor materials can also have properties that lend themselves to applications[77]). Low temperature growth results in sub-optimal physical properties for the non-magnetic III-V components in the device, limiting their optical and electrical properties. Recent work has suggested a possible solution to this dilemma by developing atomic layer epitaxy techniques[78] that can either n or p dope the non-magnetic spacer regions in GaAs/MnAs "digital ferromagnetic heterostructures".[79]

Several "proof-of-concept" device configurations for $Ga_{1-x}Mn_xAs$ have already been demonstrated within the constraints of low temperature growth, including exchange-biased samples,[80] spin-dependent resonant tunneling diodes,[81] magnetic tunnel junctions,[82,83,84] and spin-polarized light emitting diodes ("spin LEDs").[87,85,86] Additionally, tunable wave function overlap between magnetic ions and carriers confined to quantum structures has already been demonstrated in prototypical devices.[27,87] Observations of large magneto-transport effects suggest the possibility of developing very sensitive magnetic field sensors using the "giant planar Hall effect."[88] Finally, recent theoretical proposals suggest interesting new device opportunities[5,89] in bipolar configurations that combine n-doped semiconductors with $Ga_{1-x}Mn_xAs$.[90]

Most existing device configurations that exploit the spin degree of freedom are extensions of ordinary electronic devices, with enhanced but not qualitatively new functionality. Attempts have been made to explore the potential of ferromagnetic semiconductors for new functions such as reconfigurable logic using a unipolar spin transistor in which magnetic domain walls define regions wherein carriers of opposite spin play a role analogous to that of electrons and holes in conventional transistors.[91] Although implementation remains a challenge, the first experiments inspired by such proposals are already underway in nanopatterned $Ga_{1-x}Mn_xAs$ and show striking magnetoresistance effects of fundamental interest.[92]

**5. Future prospects and outstanding issues**

While $Ga_{1-x}Mn_xAs$ is an important model system, it is not yet technologically useful, primarily because its ferromagnetism does not persist up to room temperature. Although the simplest mean-field theories do suggest that room temperature ferromagnetism in $Ga_{1-x}Mn_xAs$ would be possible if the carrier and Mn densities were large enough, they become less reliable just where they predict higher transition temperatures. Will room temperature ferromagnetism eventually be achieved in $Ga_{1-x}Mn_xAs$? Theory cannot hope to provide a definitive answer, and the question is almost impossible to answer in the negative on experimental grounds. Our understanding of the materials physics is, however, becoming more complete, and we thus are slowly running short of additional experimental parameters to vary in order to enhance the energy scale of the ferromagnetic coupling in this material system.

Happily, ongoing work on a large number of candidate diluted magnetic semiconductors may ultimately yield a material with properties superior to $Ga_{1-x}Mn_xAs$.[6-19] Some of these materials (for instance (Zn,Cr)Te) are apparently homogeneous and have the required coupling between local moments and carriers evidenced, for example, by complementary measurements of magnetization like magnetic circular dichroism. Others are structurally inhomogeneous, either inadvertently containing metallic ferromagnetic inclusions that are only revealed by very careful microscopy[9,93] or intentionally grown as an inhomogeneous mixture of a semiconductor and a ferromagnetic metal.[12,77] In yet other materials (particularly the oxides such as (Zn,Mn)O, (Sn,Fe)$O_2$ and (Sn,Co)$O_2$), there is tantalizing evidence of bulk high temperature ferromagnetism in apparently homogeneous systems.[16-18] At this point, however, it is unclear that the ferromagnetism is mediated by a dilute carrier system and therefore electrically tunable.

As the search for the ideal ferromagnetic semiconductor material continues, it is useful to attempt a definition of its target, and we propose the following criteria as a possible scorecard:

1. The material should possess ferromagnetism that is primarily induced by a low-density carrier system so that magnetic properties can be tuned over a wide range by doping or by gates. Carrier induced ferromagnetism requires coupling between itinerant electrons and local moments that can be revealed by measurements of the anomalous Hall effect and magnetic circular dichroism that verify carrier participation in the magnetic order. Carrier participation *may* not necessarily imply carrier mediation however,[94] so carrier-density dependence of magnetic properties should ultimately be explicitly demonstrated.
2. The material should have a ferromagnetic transition temperature above 500 K so that it can be used in a wide range of applications without temperature control.
3. The material's magnetic properties should be sufficiently insensitive to the specific random magnetic ion distribution to allow reliable reproducibility.
4. The mean exchange field of the free carriers should be large enough to yield large giant magnetoresistance and tunnel magnetoresistance effects.
5. The material should have strong magneto-optical effects to allow optical readout of magnetically stored information.

6. Collective magnetic damping should be weak enough so that optical and transport spin-transfer phenomena, in which quasiparticle excitations are used to manipulate the magnetization, are feasible.

Below, we modestly suggest some guidelines for the search for new ferromagnetic semiconductors that meet the above criteria, fully understanding that serendipity may trump wisdom and that (our) present wisdom may prove ephemeral, or even unwise. These rules are based on the notion that if dilute moments are to be controlled by a low-density carrier system, they must be weakly coupled to other degrees of freedom.

1. The local moment ion should have a definite valence state in the host semiconductor. When this rule is not satisfied, doping will simply change the relative weights of different charge states and will not alter the effective exchange coupling between remote moments. This rule would, for example, argue against the possibility of carrier-mediated ferromagnetism in (Ga,Mn)N.

2. The local moment orbitals should hybridize strongly with orbitals at band edges of the host semiconductor in order to induce strong exchange interactions. For transition metals in zincblende semiconductors, this rule favors hole-doped materials over electron-doped materials because *p-d* hybridization occurs at the zone-center. Host semiconductors with smaller lattice constants would be favored by this rule. The position of the d-levels relative to the valence band edge also plays a key role in controlling the strength of this hybridization.

3. Degeneracies at band edges are preferable because they allow for the simultaneous occurrence of a high-density-of states and small mass bands. The latter prevent the holes from localizing near one moment and provide the magnetic stiffness that is required for a high transition temperature. The former allow for a large band spin susceptibility which favors local spin-polarization and leads to a large mean-field critical temperature.

4. It is an advantage for the local moment to have zero orbital angular momentum. Entropy associated with non-zero angular-momentum atomic moments is normally resolved in part by spontaneous lattice distortions, which causes ferromagnetic order to compete with orbital ordering and can lead to more complex magnetic states.

Based on the above criteria, we can comment about some of the more promising new diluted magnetic semiconductor materials. For Co doped ZnO, it appears clear that the ferromagnetism is not carrier induced because there is no correlation between magnetic and transport properties. Co doped $TiO_2$ does have this correlation and may have carrier induced ferromagnetism, with oxygen vacancies playing a key role in the ferromagnetism. In (Zn,Cr)Te, ferromagnetic superexchange between Cr ions -- rather than carrier-ion exchange – likely creates ferromagnetism. The Cr ion fraction has to be large in this case because the superexchange interaction is short range. This may be an interesting material, but if -- as we suspect -- its ferromagnetism is not carrier mediated, the magnetic properties will not be substantially tunable by gates and/or co-doping. A more modest change in materials – staying closer to GaAs – may be the best strategy. The quarternary alloy $Ga_{1-x}Mn_xAs_{1-y}P_y$ holds promise because the relative positions of the d-orbital energy levels and the valence band may be modified by tuning the anion fraction. The exchange interaction should become stronger and shorter range in going from $y = 0$ to $y = 1$, suggesting that an optimal alloy composition that balances the exchange strength with its range of influence could occur at an intermediate value of $y$.

As amply demonstrated by the recent work on $Ga_{1-x}Mn_xAs$, defects will play a crucial role in all potential materials. The case of $Ga_{1-x}Mn_xAs$ is an optimistic reminder that the complex defect physics of these systems is an opportunity not a warning. The magnetic properties of any diluted magnetic semiconductor ferromagnet *can* be improved by understanding and learning to control defects. Pauli's filthy mess, we predict, will (once again) be transformed by this research into a malleable magnetoelectronic clay for tomorrow's engineers.

**Acknowledgements:** This research has been supported by DARPA, ONR, the Welch Foundation and the National Science Foundation. We are also grateful to the many collaborators and colleagues whose views and insights we imperfectly reflect.

Correspondance and requests for materials should be sent to: AHM (macd@physics.utexas.edu), PS (schiffer@phys.psu.edu), or NS (nsamarth@phys.psu.edu)

Figure 1

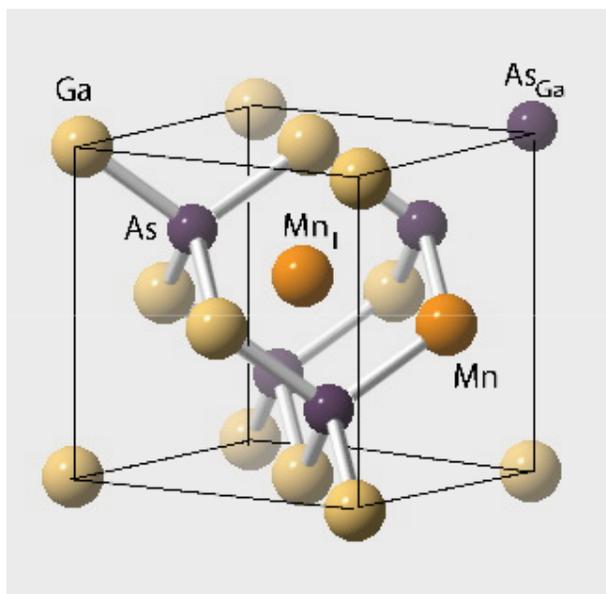

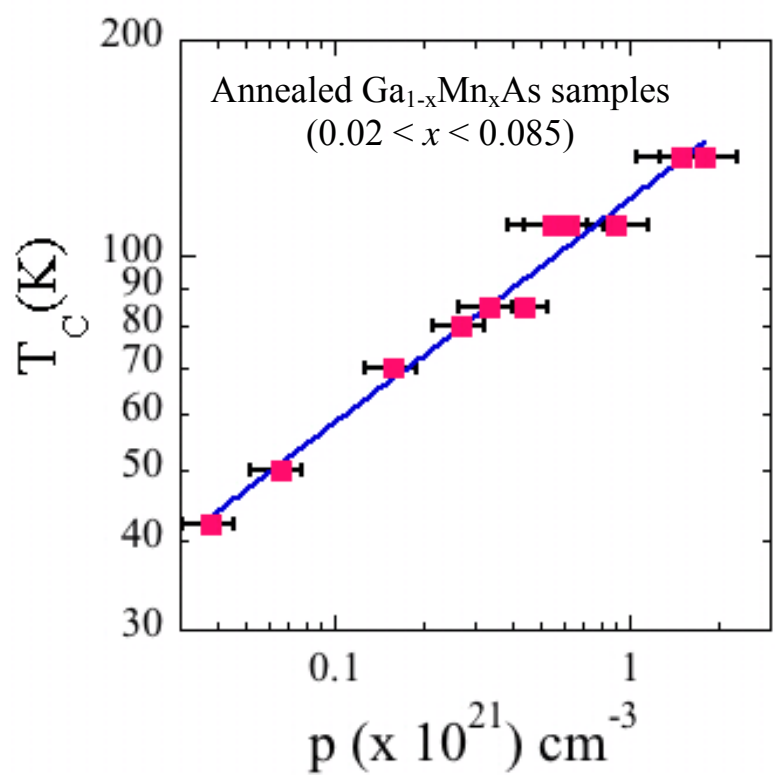

**Figure 2.**

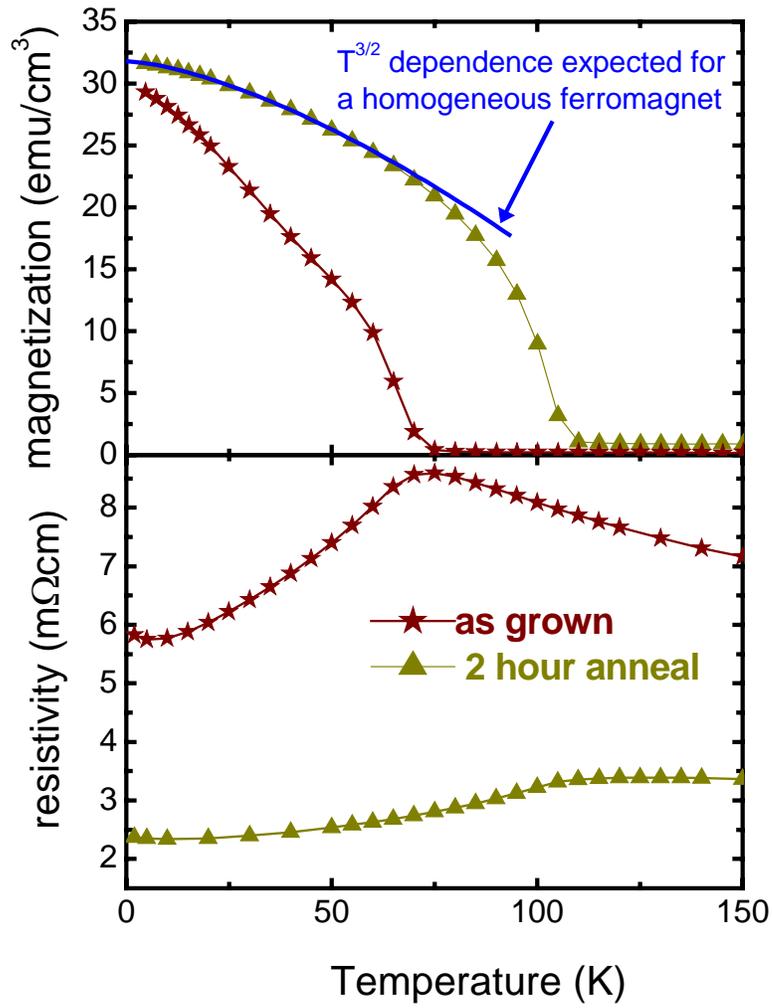

**Figure 3.**.

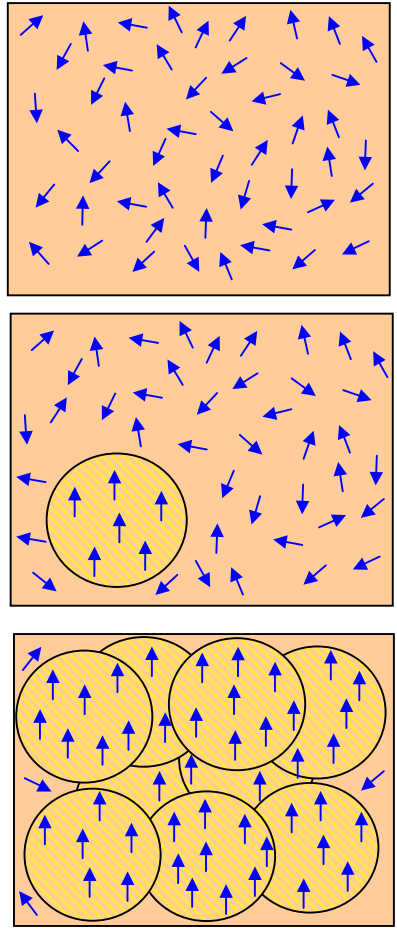

Figure 4.

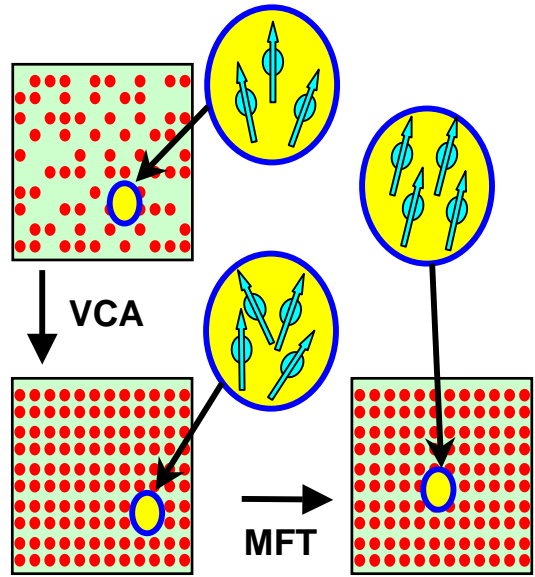

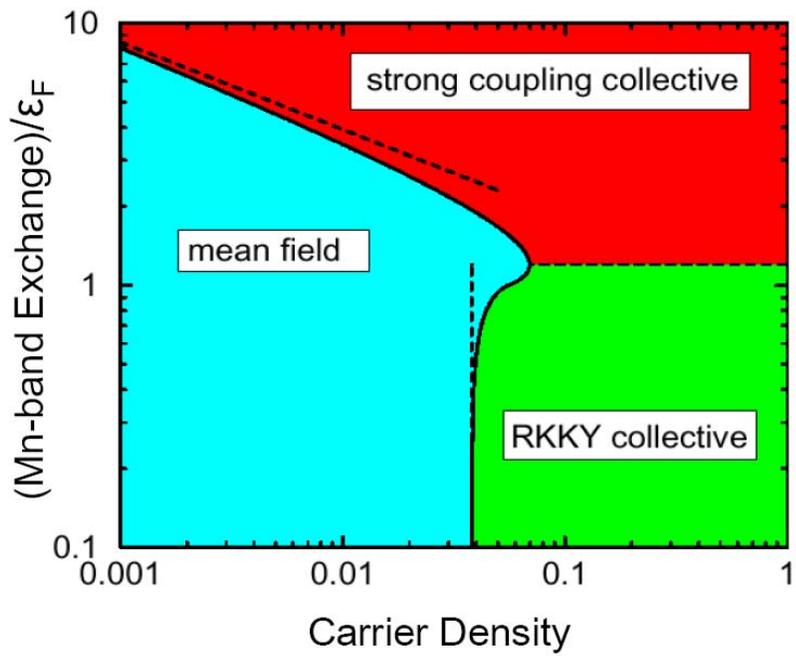

**Figure 1.** The lattice structure and transition temperature trend in $Ga_{1-x}Mn_xAs$. Top panel: Lattice of $Ga_xMn_{1-x}As$, showing a unit cell with defects: $As_{Ga}$ is an As antisite and $Mn_I$ represents a Mn interstitial. Lower panel: experimental measurements of the Curie temperature as a function of hole density in a set of consistently annealed samples of $Ga_{1-x}Mn_xAs$. The carrier densities are obtained from Raman scattering measurements, and the Curie temperature from SQUID magnetometry. The samples have a wide range of Mn content ($0.02 < x < 0.085$) and also vary in thickness (between 300 nm and 1200 nm). We note that the data show an empirical relationship $T_C \sim p^{1/3}$, apparently independent of other physical parameters such as the Mn content and sample thickness. The data are taken from Reference 29.

**Figure 2.** The temperature dependence of the magnetization and resistivity of $Ga_{0.083}Mn_{0.917}As$. The two curves in each are for unannealed (as grown) and annealed samples, and they reveal the striking physical changes wrought by annealing (increased $T_C$ and conductivity, and conventional behavior of the temperature dependent magnetization).

**Figure 3.** Pictorial explanation of carrier-induced ferromagnetism in semiconductors. In most (II,Mn)VI semiconductors, substitutional Mn impurities have $Mn^{++}$ valence states. The half-filled d shells of these ions have spin S=5/2 Hund's rule spins. Unless the semiconductors are doped with carriers, the interactions between most of these moments are weak, and they tend not to order. Instead, the spin-orientations tend to be random as indicated by the arrows in the upper panel. When they have a $Mn^{++}$ valence state, as they do in (III,Mn)V semiconductors, substitutional Mn impurities act as acceptors which localize a valence band hole, suggested by the shaded region in the middle panel. The valence band spin density near each Mn impurity tends have an orientation opposite to that of the Mn ion, because of hybridization and level repulsion between orbitals of the same spin. The valence band spins tend to spread out because of the kinetic or band energy cost of localization in quantum physics and therefore tend to interact with a number of Mn ions. The number of valence band holes can differ from the number of Mn impurities because of compensating defects, including interstitial Mn ions, that are present in great numbers in the materials unless special care is taken. If the holes are sufficiently dense and they are sufficiently spread out, they can mediate effective interactions between nearly all substitutional Mn moments, as suggested by the bottom panel. This is the situation in which homogeneous ferromagnetism with relatively high transition temperatures occurs.

**Figure 4.** Theory of Carrier-Induced Ferromagnetism in Semiconductors. Top Panel: Schematic illustration of two simplifying approximations that are valid for systems with long-range coupling between Mn spins. The first approximation replaces the random distribution of Mn ions on the host lattice by a continuum with the same density, thereby replacing the random alloy by an effective perfect crystal. (This is the virtual crystal approximation or VCA). The second approximation assumes that fluctuations in the relative orientations of Mn ions in different parts of the system do not have a large influence on the critical temperature. (This is mean field theory or MFT). Bottom Panel: A schematic phase diagram for carrier induced ferromagnetism in diluted magnetic semiconductors as a function of the exchange coupling strength (abscissa) relative to the band Fermi energy ($\varepsilon_F$) and the carrier concentration (ordinate) relative to the Mn concentration. The virtual crystal approximation will fail when the exchange interaction between Mn and band spins is too strong or the band electron masses are too small (the strong coupling region), localizing the band response to an individual spin. The mean field theory will fail in the strong coupling region and when the sign of the carrier-mediated interaction fluctuates rapidly in sign (the RKKY region). These oscillations are suppressed when the carriers are in the valence band. The largest ferromagnetic transition temperatures tend to occur near where the three regions intersect.